\documentclass[twocolumn]{aastex631}

\usepackage{xspace, color, comment}
\usepackage{amsmath, amssymb, booktabs}
\newcommand{\name}{SN~2023ufx\xspace}

\newcommand{\Msun}{\hbox{\ensuremath{\rm M_\odot}}\xspace}
\newcommand{\Zsun}{\hbox{\ensuremath{\rm Z_\odot}}\xspace}

\newcommand{\Mdot}{\hbox{\ensuremath{\dot{M}}}\xspace}

\newcommand{\Ha}{H$\alpha$\xspace}
\newcommand{\Hb}{H$\beta$\xspace}

\defcitealias{scat-spec}{269}
\defcitealias{disc}{268}
\defcitealias{class}{278}
\defcitealias{redshift}{292}

\shorttitle{Late Nebular Spectra of Metal-poor SN II 2023ufx}
\shortauthors{E. C. Jennerjahn et al.}

\begin{document}

\title{Nebular Spectra of the Extremely Metal-Poor SN II 2023ufx Over a Year After Explosion: \\ A Massive Progenitor and Unique Circumstellar Environment}

\author[0009-0009-4230-8664]{Evan Jennerjahn}
\affiliation{Center for Cosmology and AstroParticle Physics, 191 W Woodruff Ave, Columbus, OH 43210}
\affiliation{Department of Astronomy, The Ohio State University, 140 W 18th Ave, Columbus, OH 43210}

\correspondingauthor{Evan Jennerjahn}
\email{jennerjahn.3@osu.edu}

\author[0000-0002-2471-8442]{Michael A. Tucker}
\altaffiliation{CCAPP Fellow}
\affiliation{Center for Cosmology and AstroParticle Physics, 191 W Woodruff Ave, Columbus, OH 43210}
\affiliation{Department of Astronomy, The Ohio State University, 140 W 18th Ave, Columbus, OH 43210}

\author[0000-0001-6017-2961]{Christopher S. Kochanek}
\affiliation{Center for Cosmology and AstroParticle Physics, 191 W Woodruff Ave, Columbus, OH 43210}
\affiliation{Department of Astronomy, The Ohio State University, 140 W 18th Ave, Columbus, OH 43210}




\author[0000-0003-0599-8407]{Luc Dessart}
\affiliation{Institut d’Astrophysique de Paris, CNRS-Sorbonne Université, 98 bis boulevard Arago, F-75014 Paris, France}
\affiliation{French-Chilean Laboratory for Astronomy, IRL 3386, CNRS, Instituto de Astrofísica,\\Pontificia Universidad Católica de Chile, Casilla 306, Santiago, Chile}

\author[0000-0003-3953-9532]{Willem~B.~Hoogendam}
\altaffiliation{NSF Graduate Research Fellow}
\affiliation{Institute for Astronomy, University of Hawai`i, 2680 Woodlawn Drive, Honolulu, HI 96822, USA}


\author[0000-0003-4631-1149]{Benjamin J. Shappee}
\affiliation{Institute for Astronomy, University of Hawai`i, 2680 Woodlawn Drive, Honolulu, HI 96822, USA}

\begin{abstract}
We present new observations of the metal-poor ($< 0.1$~\Zsun) Type II supernova 2023ufx and its host galaxy. The deep nebular spectrum, obtained roughly a year after explosion, has a triple-peaked [\ion{O}{1}] emission profile suggesting an asymmetric explosion. Comparison to nebular spectra models suggests a zero-age main sequence mass of $ M_{\rm ZAMS} \sim 25-35$~\Msun, which is supported by the low [\ion{Ca}{2}]/[\ion{O}{1}] emission-line ratio of $\approx0.4$. The diminishing, broad, boxy \Ha emission and flattening of the light curve in the late ($\sim2$ yr) photometry suggest a complex mass-loss history in the centuries to millennia before explosion. New optical and near-infrared imaging of the host galaxy confirms that it is a dwarf, with a stellar mass of $10^{6.6\pm0.1}$ \Msun and a SFR of $10^{-2.5\pm0.1}$ \Msun/yr. Results from both SED fitting and galaxy stellar mass-metallicity scaling relations all lead to an environmental metallicity estimate of 0.02--0.07~\Zsun. Taken together, these observations confirm that \name is the explosion of a very metal-poor, heavily stripped red supergiant.

\end{abstract}



\section{Introduction} \label{sec:intro}

Massive stars ($\gtrsim8M_\odot$) often end their lives as core-collapse supernovae (CCSNe). The amount of hydrogen remaining when the explosion occurs determines their spectral classification and results in varying explosion properties. Hydrogen-rich CCSNe are dubbed Type II SNe (see \citealp{ArcaviReview} for a review), while those without are considered stripped-envelope SNe (SESNe, see \citealp{HpoorReview} for a review). The hydrogen envelopes of SESN progenitors are removed by either winds or binary interactions.

The strength of stellar winds, and thus the mass-loss rate, is determined, in part, by the progenitor metallicity. The mass-loss rate, $\dot{M}$, is usually taken to be proportional to the star's metallicity with $\dot{M}\propto Z^{\alpha}$ where $\alpha$ ranges from $\sim0.5-0.9$ \citep{alpha1,alpha2,alpha3}. Modern models predict that progenitors with lower-metallicity have lower mass-loss rates, resulting in a larger final hydrogen envelope mass at fixed M$_{\rm ZAMS}$.

This metallicity dependence of massive-star winds should manifest as an evolution of SN properties across cosmic time, as the mean cosmic metallicity increases with time \citep{highzlowZ1, highzlowZ2, highzlowZ3}. Thus, CCSNe that explode at high redshift should generally come from lower-metallicity progenitors than those that explode nearby. 


The only way to obtain good S/N spectra of these high-redshift events is to use gravitational lensing, which relies on a chance alignment, and space telescopes. As a result, there are only a few spectroscopically confirmed, lensed Type II SNe at $z>1$: SN Refsdal at $z\sim1.49$ \citep{Refsdal1, Refsdal2}, another at $z\sim2.93$ \citep{C22}, SN Eos at $z\sim5.13$ \citep{Eos}, and SN~2025mkn at $z\sim1.37$ \citep{2025mkn}. There is also a strongly lensed Type I superluminous supernova (SLSN), SN~2025wny, at $z\sim2.01$ \citep{2025wny}. The rarity of these events, and the costs of observing them, hinders in-depth study of such SNe.

Alternatively, we can use CCSNe exploding in local metal-poor dwarf galaxies to study the properties of the CCSNe expected at higher redshifts in detail. However, this is difficult for its own reasons. Most CCSNe in the local Universe occur in the more luminous and more metal-rich galaxies, which dominate the total star formation rate. There are currently only 3 candidate metal-poor CCSNe with metallicity $\lesssim 0.1Z_\odot$ and nebular-phase spectra: 2015bs \citep{anderson2018}, 2017ivv \citep{2017ivvGut}, and \name \citep{Tucker2024, Ravi2025}. This does not consider GRB SNe or SLSNe, which likely do not have RSG progenitors \citep{GRBProgen, SLSNProgen}. The 3 candidate metal-poor CCSNe exhibit some differences from their more metal-rich counterparts, such as higher luminosities, faster ejecta velocities, and slightly shorter plateaus, but the current sample size is very limited. 


Here we provide new observations of the most recent of these, \name. \citet{Tucker2024} and \citet{Ravi2025} examined the explosion with early spectra starting just a few days after detection to $\sim 200$ days after explosion. Both studies agree that the progenitor of \name must have been massive, and ended its life with a thin H envelope ($\sim1$\Msun), in order to explain its short ($\sim20$ days) plateau phase. They both found unusually high $^{56}\rm Ni$ mass estimates, with \citet{Tucker2024} finding $(0.13\pm0.04)$~\Msun and \citet{Ravi2025} finding $(0.14\pm0.02)$~\Msun. The two studies disagree about the presence of circumstellar medium (CSM) interaction in explaining the observed properties of \name. \citet{Tucker2024} argues it could explain the boxy emission profiles of [\ion{O}{1}] and \Ha, but that the lack of a blue continuum and the faintness of its Swift UVW2 flux are convincing arguments against it. \citet{Ravi2025}, on the other hand, argue that a CSM of mass $\sim0.09$~\Msun is necessary to explain its early ($\lesssim10$ days) light curve.

Here, we present new nebular spectra of \name,  taken roughly a year after explosion, as well as new deep photometry of its host galaxy. We use this photometry to better constrain the host properties in \S \ref{sec:host}, showing that it is among the most metal-poor CCSNe yet discovered. The new spectrum is analyzed in detail, and we explain several of the unique features we see in \S \ref{sec:nebspec}.

\section{Host-Galaxy Properties}\label{sec:host}

We obtained optical and near-IR images of the host galaxy with the Large Binocular Cameras (LBC) \citep{LBC1,LBC2} and LUCI \citep{LUCI} instruments on the Large Binocular Telescope (LBT). The images were reduced using standard procedures, including bias and dark subtraction, and flat-field corrections. We use Sloan Digital Sky Survey (SDSS; \citealp{sdss1}) $ugriz$ photometry and the \citet{Jester2005} conversions to photometrically calibrate the $UBVRI$ imaging, and Two Micron All-Sky Survey (2MASS; \citealp{2MASS}) photometry of stars for the JHK bands. We measured the host-galaxy magnitudes using a 4$\arcsec$ aperture centered on the host. 

Additionally, we attempted to measure the magnitude of \name in the LBC data using a 0\farcs8 aperture centered on the supernova and obtained $\gtrsim3\sigma$ detections in the $B$ and $R$ LBC filters. Even in these bands, the SN flux is $\sim2\%$ of the host flux, and thus does not significantly affect the host-galaxy photometry.

\begin{figure*}
    \centering
    \includegraphics[width=\linewidth]{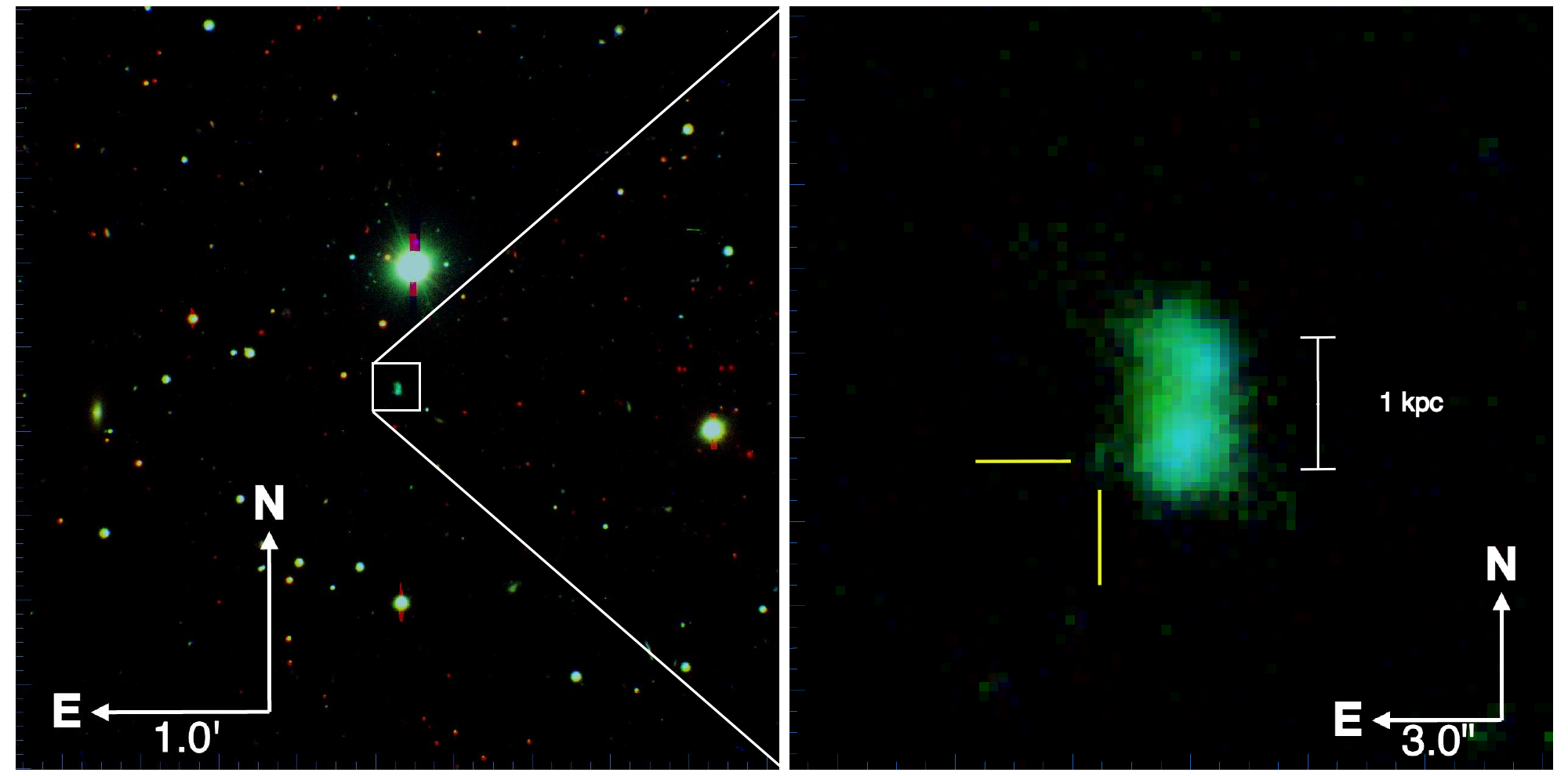}
    \caption{$UBI$ image of the host galaxy of \name with bright saturated star trails masked out. Blue is $U$, Green is $B$, and Red is $I$. A crosshair in the right panel marks the location of the supernova.}
    \label{fig:Host}
\end{figure*}

Fig \ref{fig:Host} shows a color image of the host from the new LBT data. We supplemented the new data with archival photometry from the Galaxy Evolution Explorer (GALEX; \citealp{GALEX,morrissey2007}),  SDSS; \citealp{sdss1}, Dark Energy Survey (DES; \citealp{DES1}), and Panoramic Survey Telescope and Rapid Response System (Pan-STARRS; \citealp{Panstarrs1}). All host-galaxy photometry information is provided in Table \ref{tab:MagInfo}, where the archival magnitudes were obtained using \textsc{Blast} \citep{Blast}.

\begin{figure}
    \centering
    \includegraphics[width=\linewidth]{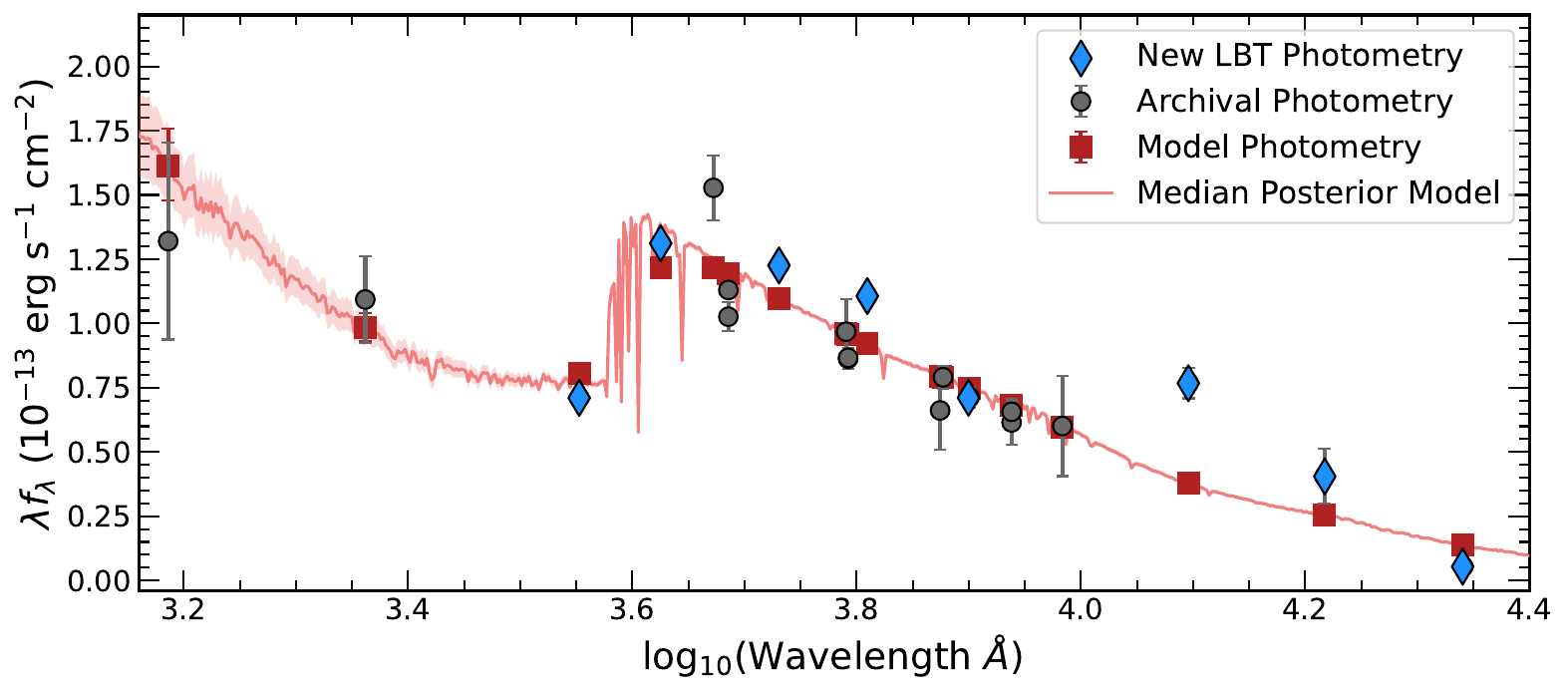}
    \caption{SED of the host of \name using archival photometry (grey circles) paired with new $UBVRI$ and $JHK$ photometry from the LBT (blue diamonds). The red line is the median posterior SED model. The shaded band shows the 1$\sigma$ scatter in the models.}
    \label{fig:SED}
\end{figure}

We fit the SED using \textsc{BAGPIPES} \citep{Bagpipes}, assuming a delayed exponential star formation history and the dust attenuation model from \citet{SalimDust} with a $\delta$ value of $-0.45$ and a 2175 \AA \xspace bump strength of 1 (roughly SMC values). The fit is shown in Fig. \ref{fig:SED}, and the parameters are listed in Table \ref{tab:HostInfo}. We find a stellar mass of $10^{6.6\pm0.1}$\Msun and a metallicity of $(0.02\pm0.01)$\Zsun. We also tried an exponential star formation rate model instead of a delayed one, and found little change in the results.

\begin{table}[bt]
    \begin{center}       
    \begin{tabular}{ccc}
    \hline
    \hline
    \rule[-1ex]{0pt}{3.5ex} Facility & Filter & Magnitude \\
    \hline
    \rule[-1ex]{0pt}{3.5ex} LBT/LBC & $U$  & $20.83\pm0.05$ \\
    \rule[-1ex]{0pt}{3.5ex} LBT/LBC & $B$  & $20.87\pm0.02$ \\
    \rule[-1ex]{0pt}{3.5ex} LBT/LBC & $V$  & $20.59\pm0.03$ \\
    \rule[-1ex]{0pt}{3.5ex} LBT/LBC & $R$  & $20.23\pm0.03$ \\
    \rule[-1ex]{0pt}{3.5ex} LBT/LBC & $I$  & $20.36\pm0.06$ \\
    \rule[-1ex]{0pt}{3.5ex} LBT/LUCI & $J$  & $19.25\pm0.08$ \\
    \rule[-1ex]{0pt}{3.5ex} LBT/LUCI & $H$  & $19.21\pm0.29$ \\
    \rule[-1ex]{0pt}{3.5ex} LBT/LUCI & $K$  & $20.57\pm0.24$ \\
    \rule[-1ex]{0pt}{3.5ex} Pan-STARRS & $g$  & $20.85\pm0.06$ \\
    \rule[-1ex]{0pt}{3.5ex} Pan-STARRS & $r$  & $20.77\pm0.05$ \\
    \rule[-1ex]{0pt}{3.5ex} Pan-STARRS & $i$  & $20.65\pm0.06$ \\
    \rule[-1ex]{0pt}{3.5ex} Pan-STARRS & $z$  & $20.78\pm0.15$ \\
    \rule[-1ex]{0pt}{3.5ex} Pan-STARRS & $y$  & $20.69\pm0.35$ \\
    \rule[-1ex]{0pt}{3.5ex} SDSS & $g$  & $20.45\pm0.09$ \\
    \rule[-1ex]{0pt}{3.5ex} SDSS & $r$  & $20.65\pm0.14$ \\
    \rule[-1ex]{0pt}{3.5ex} SDSS & $i$  & $20.85\pm0.25$ \\
    \rule[-1ex]{0pt}{3.5ex} DES & $g$  & $20.75\pm0.02$ \\
    \rule[-1ex]{0pt}{3.5ex} DES & $r$  & $20.77\pm0.04$ \\
    \rule[-1ex]{0pt}{3.5ex} DES & $z$  & $20.70\pm0.08$ \\
    \rule[-1ex]{0pt}{3.5ex} GALEX & FUV  & $21.82\pm0.32$ \\
    \rule[-1ex]{0pt}{3.5ex} GALEX & NUV  & $21.59\pm0.17$ \\
    \hline 
    \hline
    \end{tabular}
    \end{center}
    \caption{Magnitudes used for SED model of the host of \name. The archival magnitudes were obtained using \textsc{Blast} \citep{Blast}} 
    \label{tab:MagInfo}
\end{table} 

\begin{table}[bt]
    \begin{center}       
    \begin{tabular}{cc}
    \hline
    \hline
    \rule[-1ex]{0pt}{3.5ex} Parameter  & Value \\
    \hline
    \rule[-1ex]{0pt}{3.5ex} log$_{10}$(Mass [M$_\odot$])  & $6.6\pm0.1$ \\
    \rule[-1ex]{0pt}{3.5ex} log$_{10}$(SFR [M$_\odot$ yr$^{-1}$])  & $-2.5\pm0.1$ \\
    \rule[-1ex]{0pt}{3.5ex} log$_{10}$(sSFR [yr$^{-1}$])  & $-9.1\pm0.1$ \\
    \rule[-1ex]{0pt}{3.5ex} Mass Weighted Age (Gyr) & $0.6\pm0.1$ \\
    \rule[-1ex]{0pt}{3.5ex} $\tau$ (Gyr) & $0.34\pm0.05$ \\
    \rule[-1ex]{0pt}{3.5ex} A$_{\rm v}$ (mag)  & $0.05\pm0.04$ \\
    \rule[-1ex]{0pt}{3.5ex} Metallicity from SED [Z$_\odot$]  & $0.02\pm0.01$ \\
    \rule[-1ex]{0pt}{3.5ex} Metallicity using M-Z relation [Z$_\odot$]  & $0.07^{+0.07}_{-0.04}$ \\
    \hline 
    \hline
    \end{tabular}
    \end{center}
    \caption{Properties of the host galaxy of \name from the SED fit using new LBT data supplemented with archival photometry.} 
    \label{tab:HostInfo}
\end{table} 

\begin{figure*}
    \centering
    \includegraphics[width=\linewidth]{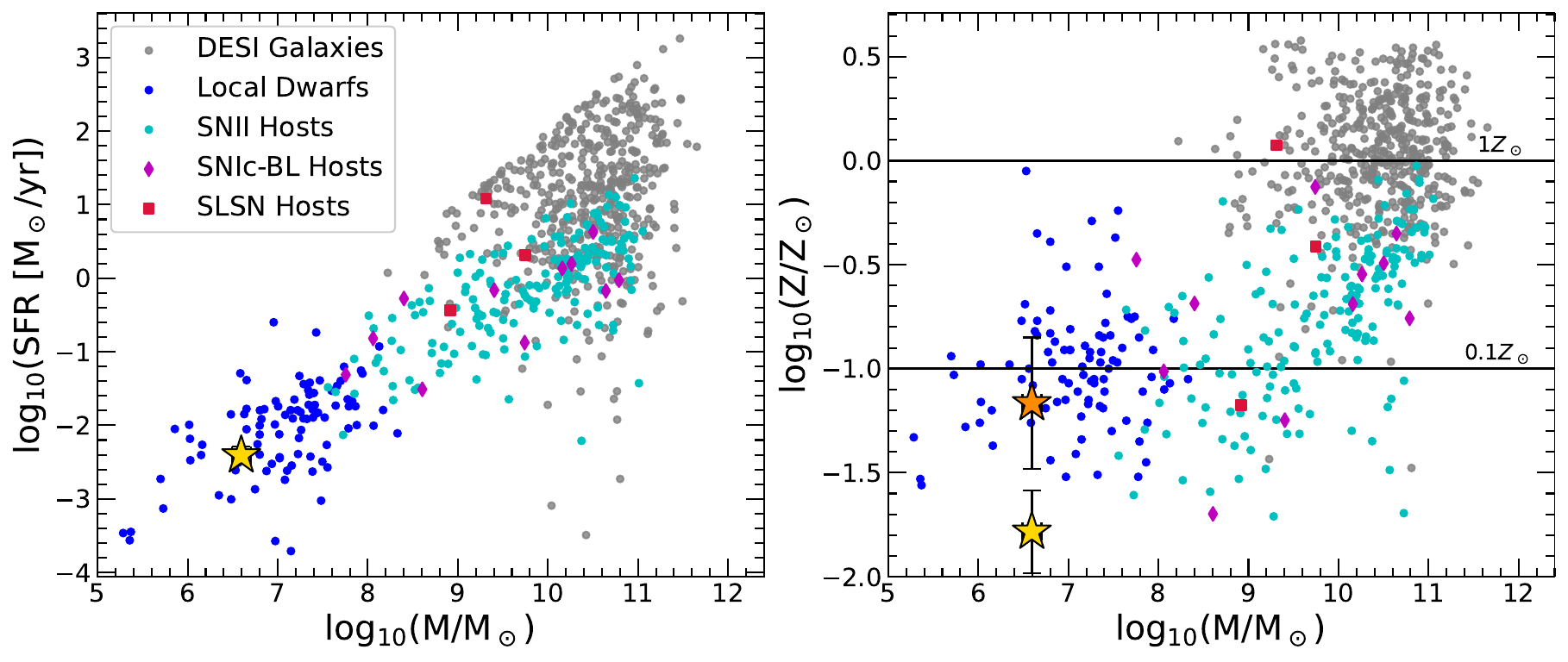}
    \caption{The galaxy M$_{\rm star}$-SFR (left) and M$_{\rm star}$-metallicity (right) planes, including the host-galaxy of \name. The mass and SFR for the host of \name come from the SED fit using new LBT data (gold). The M-Z plane (right) shows the metallicity derived from the SED model (gold) in addition to the one derived with the M-Z relation of \citet{BergMZ} using the SED estimate of the stellar mass (dark orange).}
    \label{fig:DESI}
\end{figure*}

In SED models, the metallicity estimate is usually degenerate with other properties such as age and dust \citep{SEDMetalDegen}, so we also used the \citet{BergMZ} mass-metallicity (M-Z) relation to estimate it, finding Z of $0.07^{+0.07}_{-0.04}$~\Zsun. Both metallicity estimates are shown in the right panel of Fig. \ref{fig:DESI}, where we compare the host of \name to other galaxies. For comparison, we show estimates for DESI galaxies from \citet{Desi_Zou}, where we require log(sSFR)$<-8$ to prevent quasar contamination, and low-metallicity blue compact dwarf galaxies from \citet{hsyu2018}. We also show host properties of CCSNe from the SCAT DR1 \citep{SCATDR1} derived with the \textsc{Blast} SED fitting infrastructure \citep{Blast}. 

\citet{pessi2023b} compiled metallicity estimates of a homogeneous sample of ASAS-SN CCSNe using integral-field spectroscopy and strong-line metallicity calibrations (e.g., \citealp{D16}). Their catalogs give a minimum environmental metallicity of $\sim0.09$~\Zsun for a subset of 81 H-rich (II, IIP, IIb) SNe within a redshift of 0.02. This is higher than our estimates for the host of \name, although within the uncertainty of our value using the M-Z relation of \citet{BergMZ}.

\section{Nebular Spectra of \name}\label{sec:nebspec}

New nebular spectra were obtained with the Keck Cosmic Web Imager (KCWI; \citealp{kcwi_morrissey}) on the Keck II telescope at a phase of +184d, and with the Multi-Object Double Spectrographs (MODS; \citealp{pogge2010}) on the LBT at phases of +361d, +364d, and +385d. The MODS spectra were combined into a final spectrum with a mean phase after explosion of +370d. Spectra were reduced with \textsc{pypeit} \citep{pypeit_prochaska} using standard reduction and flux calibration techniques. We use the r-band acquisition images and an $A_V$ of 0.124 mag \citep{Tucker2024} to flux-calibrate the combined +370d spectrum.

\begin{figure*}
    \centering
    \includegraphics[width=\linewidth]{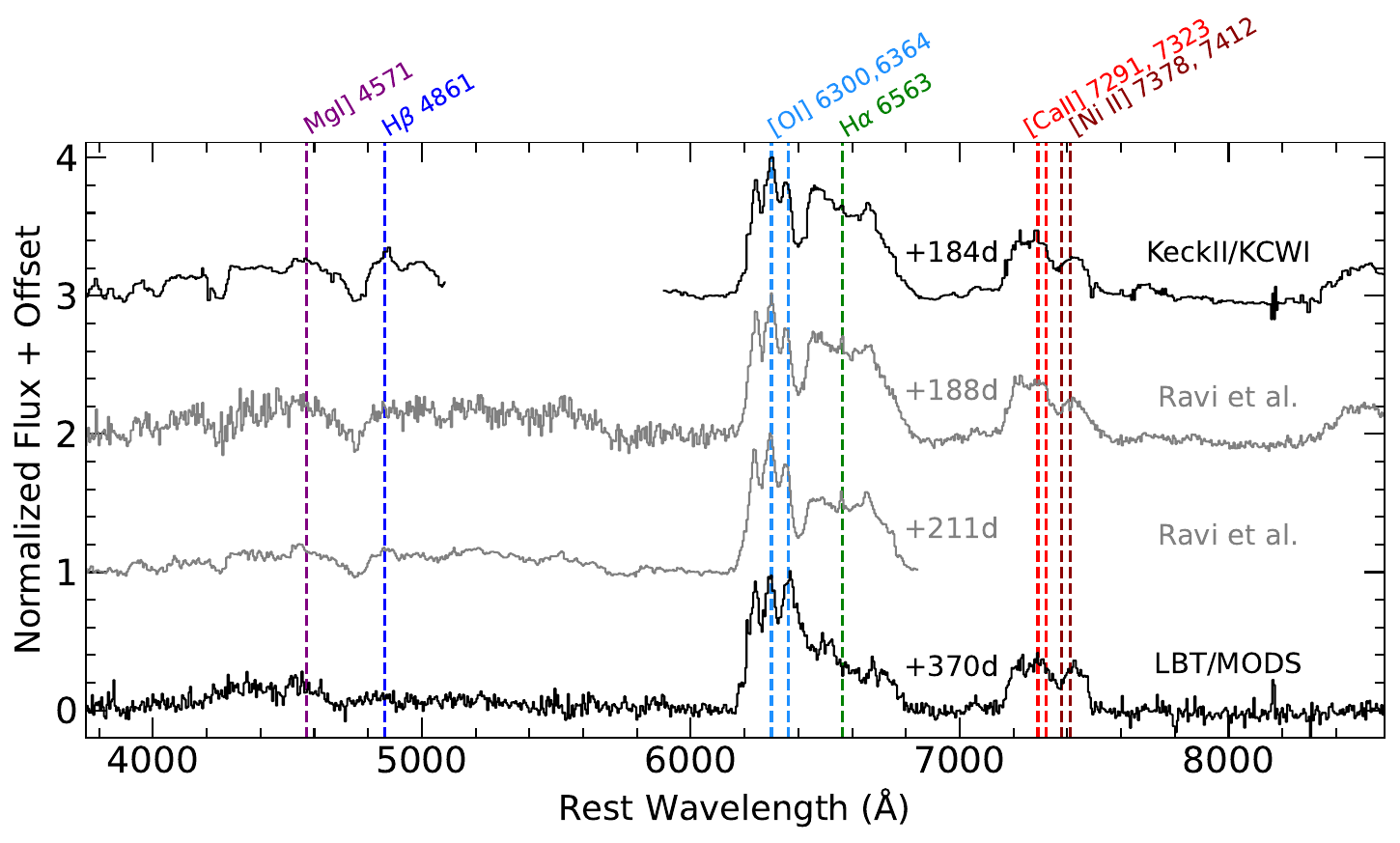}
    \caption{Nebular evolution of \name. Spectra from this work are shown in black, and the intervening spectra from \citet{Ravi2025} are shown in gray. Vertical lines denote important features discussed in \S\ref{sec:nebspec}.}
    \label{fig:NebSpec}
\end{figure*}

\begin{figure*}
    \centering
    \includegraphics[width=\linewidth]{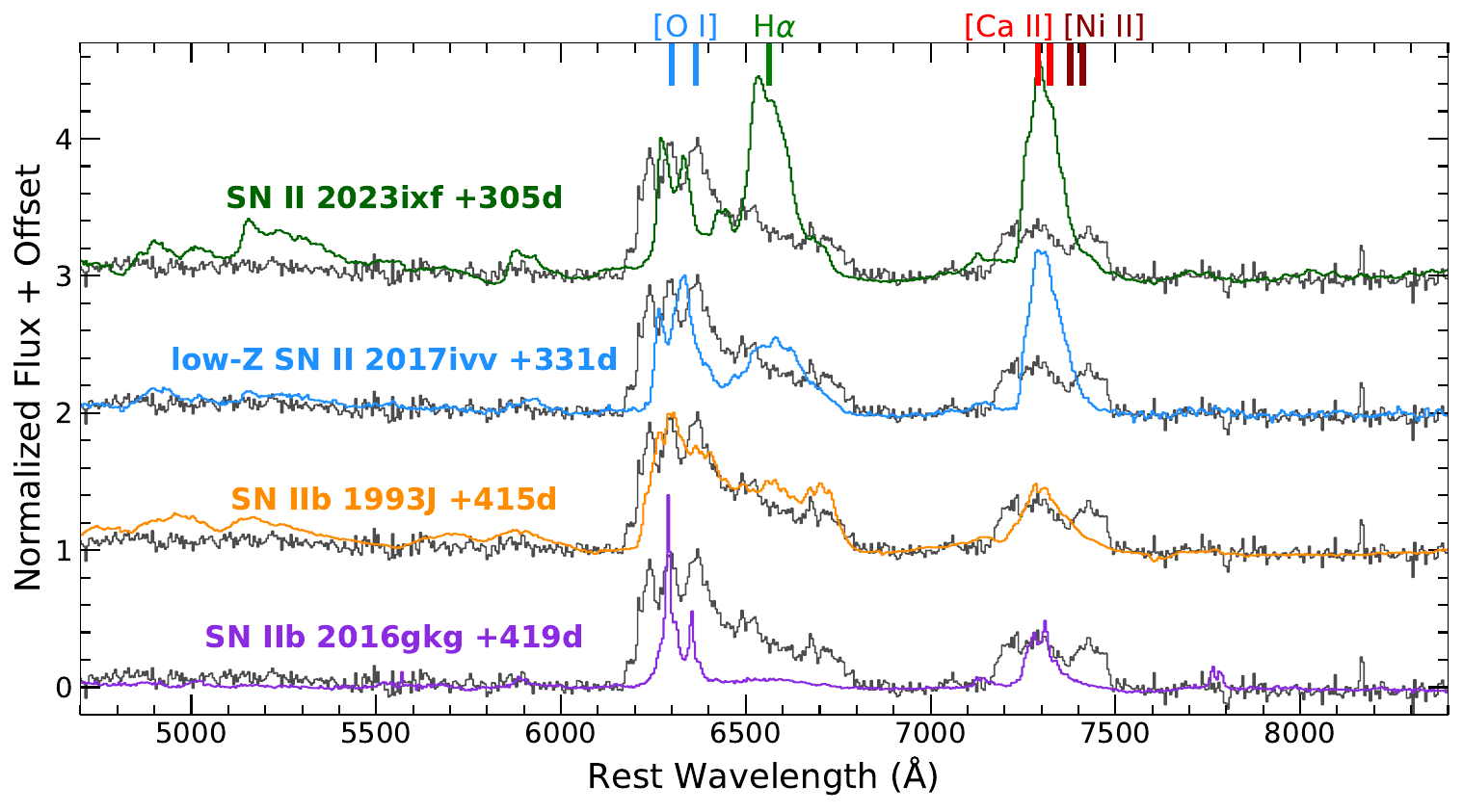}
    \caption{A comparison of the +370d nebular spectrum of \name (black) to other well-observed SNe II at similar phases (see \S \ref{sec:nebspec}). SN 2023ixf (top; \citealp{ixfspec}) is a normal CCSN, SN~2017ivv (2nd from top; \citealp{2017ivvGut}) is another low-metallicity SN II, SN~1993J (2nd from bottom; \citealp{93Jspeccite}) is a Type IIb whose spectrum is the closest match we were able to find to \name, and SN~2016gkg (bottom; \citealp{SN16gkg}) is an example of an asymmetric SN IIb.} 
    \label{fig:CompSNe}
\end{figure*}

The nebular spectra of \name are shown in Fig. \ref{fig:NebSpec}, and we compare the +370d spectrum to other well-observed CCSNe in Fig. \ref{fig:CompSNe}. We first compare to the normal SN II 2023ixf \citep{ixfGeneralCite}, which shows a poor match to the nebular spectrum of \name. Thus, we also compare it to other unusual H-rich CCSNe. SN~2017ivv is chosen because it is another of the small sample of very metal-poor CCSNe \citep{2017ivvGut}. SN~1993J is perhaps the best studied example of a Type IIb SN, has convincing evidence of a binary companion \citep{93JBinar1,93JBinar2,93JComp,93JCompHST}, and its nebular spectrum is the closest match we could find for \name. SN~2016gkg is another SN IIb with evidence for explosion asymmetry \citep{SN16gkg}.

Intriguingly, the best match is to the nearby SN~1993J, which exploded in the luminous, metal-rich galaxy M81. Yet even this match is imperfect, as none of the comparison objects show the distinctly triple-peaked [\ion{O}{1}] profile nor have similar structures near 7300\xspace\AA. This is discussed further in \S \ref{sec:Disc}.

\subsection{Progenitor Mass}\label{subsec:ProMass}

\begin{figure*}
    \centering
    \includegraphics[width=\linewidth]{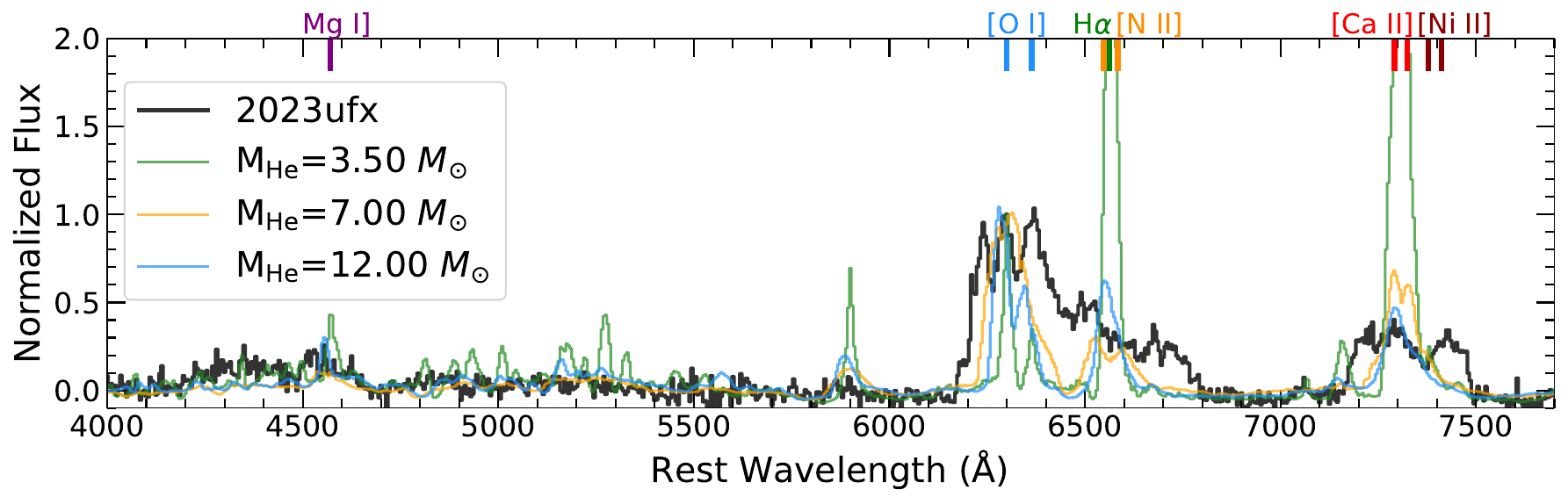}
    \caption{A comparison of the +370d spectrum of \name to the SESN nebular spectra models of \citet{DessartStripped} at +360d, which assume no rotation and solar metallicity. The helium core masses of 3.50, 7.00, and 12.00\Msun correspond to approximate ZAMS masses of 17, 26, and 36\Msun, respectively.}
    \label{fig:CompModels}
\end{figure*}

Nebular spectra of CCSNe probe the conditions of the progenitor's helium core at the time of explosion, and are less sensitive to stripping from a companion or other mass-loss effects, thus allowing us to estimate the progenitor mass. We use two methods to estimate the progenitor ZAMS mass from the nebular spectra. First, we directly compare against the nebular models of stripped stars from \citet{DessartStripped}. These models are at solar metallicity, ignore rotation, and range in helium core (ZAMS) mass from 2.5--12.0~\Msun ($\sim14-36$~\Msun). We show comparisons between \name and the models in Fig. \ref{fig:CompModels}. The low-mass models (e.g., M$_{\rm He}$=3.50~\Msun) significantly overestimate the [\ion{Ca}{2}] emission relative to [\ion{O}{1}]. The M$_{\rm He}$=7.00~\Msun model does a better job at matching the [\ion{O}{1}] emission, but still overestimates the [\ion{Ca}{2}] emission. The M$_{\rm He}$=12.00~\Msun model is the best match to the [\ion{Ca}{2}] emission, but fails to reproduce the broad [\ion{O}{1}] flux. No model in the suite matches the sloped continuum seen around \Ha, nor the multi-peaked emission profiles around 6300 \AA \xspace and 7300 \AA. The best matches come from the higher mass progenitors, implying a high mass ($\gtrsim 25$~\Msun) progenitor for \name.

Since the models do not capture the unusual velocity profiles seen in the nebular emission lines of \name, we turn to the [\ion{Ca}{2}]/[\ion{O}{1}] flux ratio as a secondary proxy for the progenitor mass. The flux of [\ion{Ca}{2}] from the SN ejecta is relatively insensitive to the ZAMS mass of the progenitor, while the mass of the oxygen core, and thus the flux of [\ion{O}{1}], is directly dependent on it \citep{O2CaProof}. The flux ratio of these lines is considered a more robust indicator of the mass than the absolute [\ion{O}{1}]$\lambda\lambda6300,6364$ line luminosity because the ratio is not sensitive to errors in flux calibrations, distances, or extinctions \citep{RatioBetter1, jerkstrand2014}. Lower values of the [\ion{Ca}{2}]/[\ion{O}{1}] ratio indicate a larger O-core mass, and thus a more massive progenitor.

\begin{figure}
    \centering
    \includegraphics[width=\linewidth]{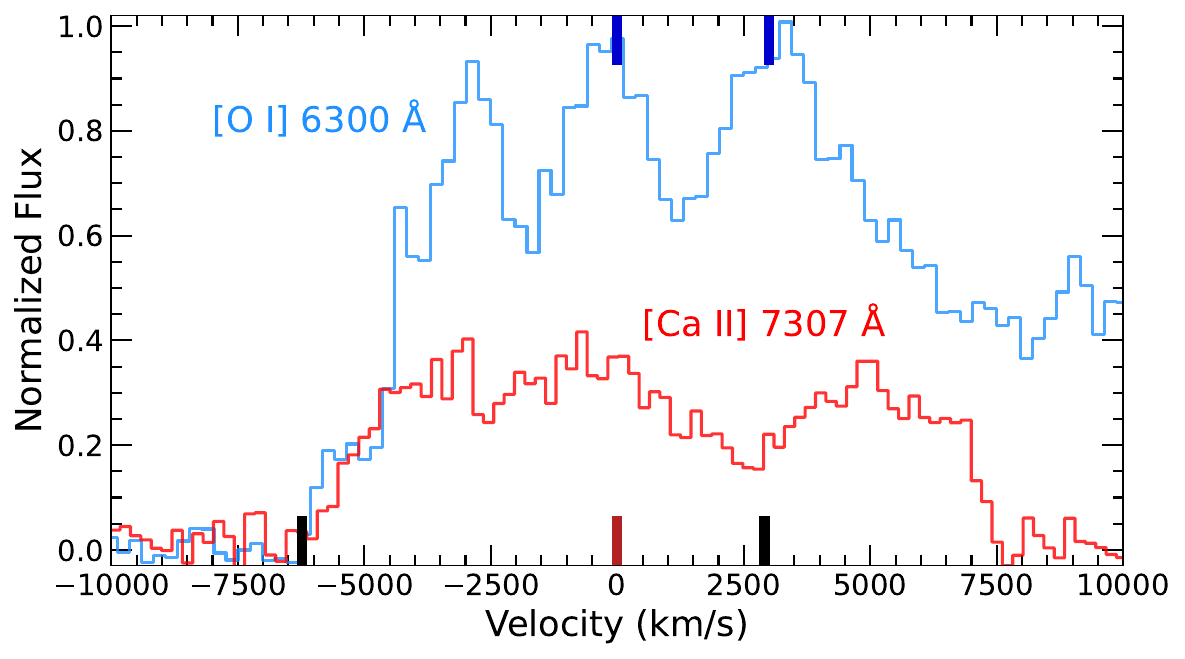}
    \caption{Velocity distributions of [\ion{O}{1}] (blue) and [\ion{Ca}{2}] (red). [\ion{O}{1}] is centered on the 6300~\AA \xspace line, and blue ticks along the top mark the rest wavelengths of the 6300 \AA \xspace and 6363 \AA \xspace doublet components. [\ion{Ca}{2}] is centered on the mean wavelength of the doublet at 7307 \AA \xspace, and black ticks mark the rest velocities of [\ion{Fe}{2}]$\lambda7155$ and [\ion{Ni}{2}]$\lambda7378$.}
    \label{fig:v_dist}
\end{figure}

Accurately measuring the emission-line fluxes is complicated by the unusual, asymmetric, velocity profiles shown in Fig. \ref{fig:v_dist}. The [\ion{O}{1}] emission profile has three peaks: a central peak at $v\approx0$ km/s, and red- and blueshifted peaks at $v\approx\pm3000$ km/s. It is unclear exactly where the [\ion{O}{1}] profile ends on the redder edge, as it becomes blended with some combination of \Ha and/or [\ion{N}{2}]. Moreover, it is unclear how much of the 7300 \AA \xspace feature can be attributed to [\ion{Ca}{2}], as [\ion{Fe}{2}] and [\ion{Ni}{2}] also have transitions in this region. [\ion{Fe}{2}] emission is unlikely given that the associated Fe blends at bluer wavelengths ($\sim4500$~\AA) are absent.

Thus, the redshifted peak of the 7300 \AA \xspace feature can either be [\ion{Ca}{2}] at $\sim5000$ km/s or [\ion{Ni}{2}] at $\sim2000$ km/s. The central and blueshifted peaks of [\ion{O}{1}] have corresponding peaks in [\ion{Ca}{2}], lending credence to their identification as [\ion{Ca}{2}]. However, the redshifted peaks are not well aligned. Thus, we consider [\ion{Ni}{2}] a more plausible identification for this feature, supported by the time-evolution of the 7300 \AA \xspace region seen in Fig. \ref{fig:NebSpec}. The central of the three bumps around 7300 \AA \xspace remains roughly equal in strength to the blueshifted bump, but both of these weaken relative to the redshifted peak at later times. 

This suggests different origins for the two features, and thus we assume the reddest feature originates from [\ion{Ni}{2}]. It does seem peculiar for [\ion{O}{1}] and [\ion{Ca}{2}] to both have blueshifted components and for [\ion{Ni}{2}] to only have a redshifted component, but we still find this to be the most likely scenario.

\begin{figure}
    \centering
    \includegraphics[width=\linewidth]{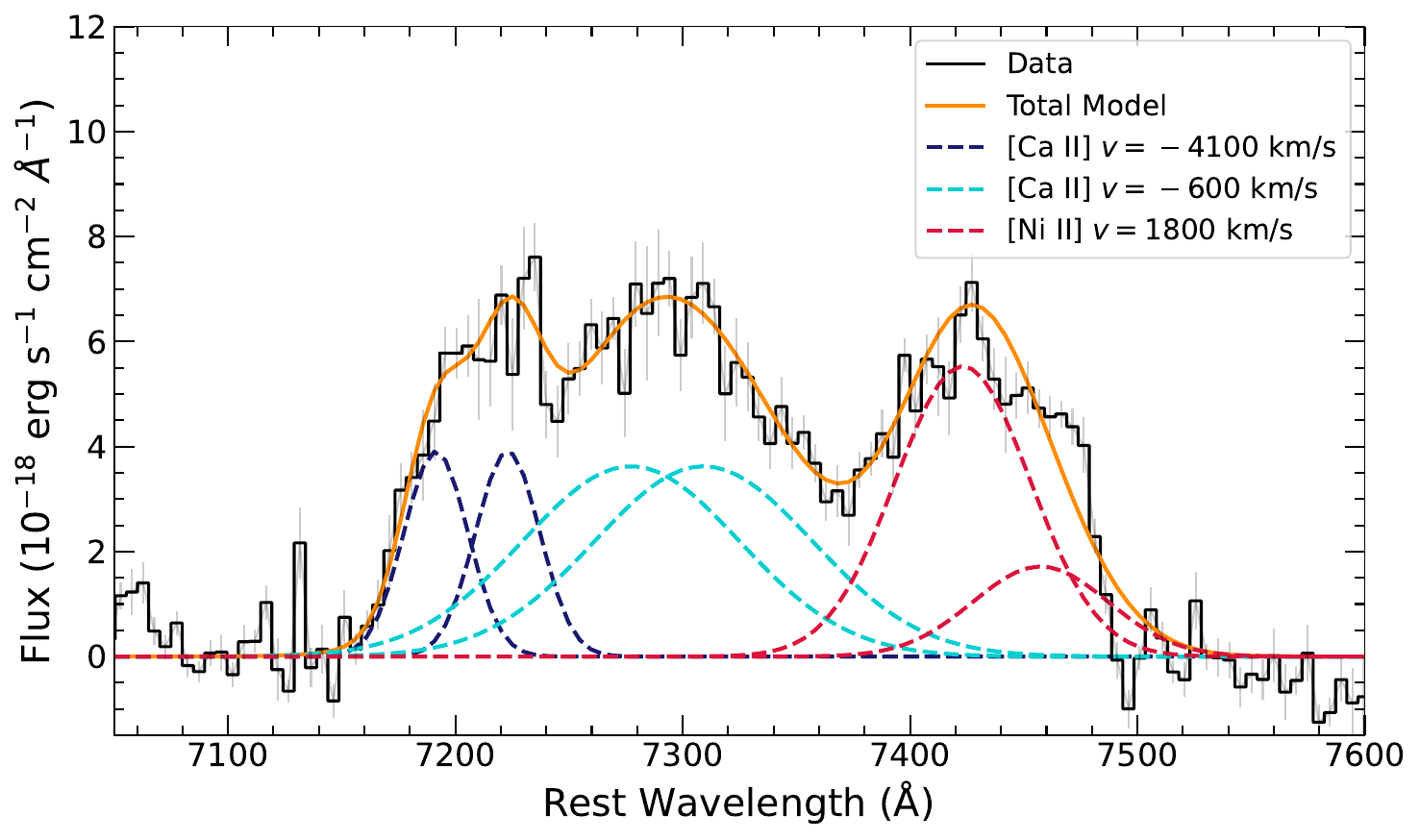}
    \caption{Fit to the 7000--7600 \AA \xspace region of the spectrum assuming the two components to the left come from blueshifted [\ion{Ca}{2}] and that the component to the right comes from redshifted [\ion{Ni}{2}]}
    \label{fig:CaNiFit}
\end{figure}

Fig. \ref{fig:CaNiFit} shows our efforts to model the 7000--7600~\AA \xspace region with a series of Gaussians. The left and central peaks are fit with two [\ion{Ca}{2}]$\lambda\lambda7291,7323$ doublets assuming different bulk velocities and FWHM values for each doublet. The right peak is fit with redshifted [\ion{Ni}{2}] where the [\ion{Ni}{2}] emission comes from a combination of redshifted [\ion{Ni}{2}]$\lambda7378$ \xspace and [\ion{Ni}{2}]$\lambda7412$ with a fixed ratio of L$_{7412}$/L$_{7378}$=0.31 \citep{NiFeJerk}, and a shared velocity and FWHM. The resulting line velocities and luminosities are given in Table \ref{tab:VelocityInfo}, where we convert to luminosity using a distance of $65.9\pm4.4$ Mpc \citep{Tucker2024}.

\begin{table*}[bt]
    \begin{center}   
    \begin{tabular}{c|c|c|c|c}
    \hline
    \hline
    \rule[-1ex]{0pt}{3.5ex} Emission Line  & Velocity (km/s) & FWHM (km/s) & Flux (10$^{-16}$ erg/s/cm$^2$) & Luminosity (10$^{38}$ erg/s)\\
    \hline
    \rule[-1ex]{0pt}{3.5ex}  [\ion{Ca}{2}] (Comp. 1) & $-4100\pm200$ & $1400\pm200$ & $2.8\pm0.3$ & $1.4\pm0.2$\\
    \rule[-1ex]{0pt}{3.5ex}  [\ion{Ca}{2}] (Comp. 2) & $-600\pm200$ & $4600\pm300$ & $8.7\pm0.4$ & $4.5\pm0.6$\\
    \rule[-1ex]{0pt}{3.5ex}  [\ion{Ca}{2}] (Total) & ... & ... & $11.5\pm0.5$ & $6.0\pm0.8$\\
    \rule[-1ex]{0pt}{3.5ex}  [\ion{Ni}{2}] & $1800\pm200$ & $2800\pm200$ & $5.4\pm0.2$ & $2.8\pm0.4$\\
    \hline 
    \hline
    \end{tabular}
    \end{center}
    \caption{Velocity and luminosity information for the [\ion{Ca}{2}] and [\ion{Ni}{2}] emission lines in the +370d spectrum of \name.} 
    \label{tab:VelocityInfo}
\end{table*} 

The [\ion{O}{1}]$\lambda\lambda6300,6364$ regions of the spectra are unique compared to the other CCSNe we show in Fig. \ref{fig:CompSNe}, showing signs of extreme explosion asymmetry. The [\ion{O}{1}] profile appears to merge with the \Ha and/or [\ion{N}{2}] feature at $\approx6400$ \AA, preventing a reliable decomposition of the individual emission components. Instead of fitting Gaussians, we sum all of the flux between 6170--6405 \AA \xspace ($v\approx\pm6000$ km/s in Fig. \ref{fig:v_dist}) to estimate the [\ion{O}{1}] luminosity. This leads to an [\ion{O}{1}] flux of $(31.2\pm0.6)\times10^{-16}$ erg/s/cm$^2$ in the +370d spectrum, corresponding to a line luminosity of $(16.2\pm2.2)\times10^{38}$ erg/s. Combined with the [\ion{Ca}{2}] fluxes in Table \ref{tab:VelocityInfo}, our final [\ion{Ca}{2}]/[\ion{O}{1}] ratio is $0.37\pm0.02$.

This ratio falls well below the cutoff value of [\ion{Ca}{2}]/[\ion{O}{1}]=1.43 used to differentiate between Type II SNe and SESNe \citep{JerkBook, 2017ivvGut, ShortPlat}, and implies that the progenitor of \name was more massive than typical SN~II progenitors ($\sim10-20$~\Msun). 

\begin{figure*}
    \centering
    \includegraphics[width=\linewidth]{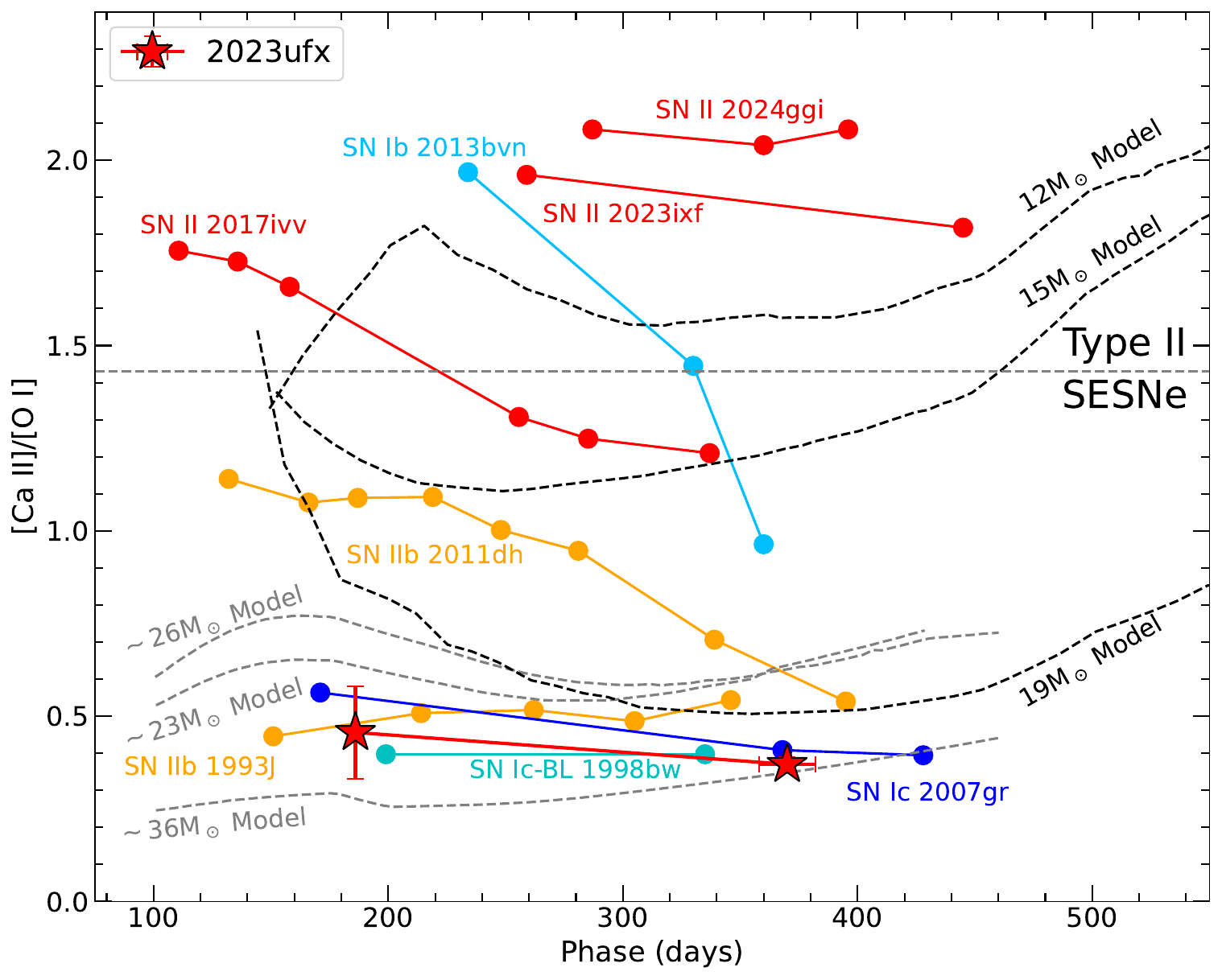}
    \caption{The [\ion{Ca}{2}]/[\ion{O}{1}] ratios of several well studied CCSNe, including \name, as well as several models. The two ratios calculated for the spectra around +180 days for \name have been averaged together. The grey dashed horizontal line represents the typical division between Type II SNe and SESNe ([\ion{Ca}{2}]/[\ion{O}{1}]=1.43). The grey dashed curves are models of SESNe from \citet{DessartStripped} evolved with wind mass-loss. The black dashed curves are the models of Type IIP SNe from \citet{JerkTimeModels}. Measurements for the observed CCSNe come from \citet{2017ivvGut}, \citet{TimeEvolve1}, \citet{TimeEvolve2}, \citet{TimeEvolve3}, and \citet{TimeEvolve4}.}
    \label{fig:TimeEvolve}
\end{figure*}

Fig. \ref{fig:TimeEvolve} places the time evolution of the [\ion{Ca}{2}]/[\ion{O}{1}] ratio of \name in context with other well-studied CCSNe and explosion models.  The ratios measured for \name are well below the typical Type II SNe, and fall between the models with helium core masses of 7~\Msun (M$_{\rm ZAMS}\approx26$~\Msun) and 12~\Msun (M$_{\rm ZAMS}\approx36$~\Msun) from \citet{DessartStripped}, which are shown in Fig. \ref{fig:CompModels}. The ratio is also close to the values for the broad-line Ic SNe (Ic-BL) 1998bw and SN Ic 2007gr. SN 1998bw has an estimated progenitor mass of 30--35~\Msun \citep{98bwRevisited}. The progenitor mass of SN 2007gr is contested, but some estimates range as high as 40~\Msun \citep{07gr40Msun}. This agrees with our previous mass estimate of $\gtrsim25$~\Msun, and leads to a better constrained estimate of $\sim$25--35~\Msun.

\subsection{CSM Environment}\label{subsec:Ha}

A noticeable feature of the spectral evolution in Fig. \ref{fig:NebSpec} is the persistent fading of the broad, boxy \Ha emission. It rivals the [\ion{O}{1}] complex at $\lesssim +200$d but has weakened considerably in the last spectrum at +370d. CSM interaction is the most plausible source of the box-like \Ha profile, which is commonly seen in interacting SNe II (e.g. \citealp{boxyCSM1,boxyCSM2,boxyCSM3}) and models of ejecta-CSM interactions (e.g. \citealp{boxyCSMmodel1}). The boxy structure reflects the top-hat line-of-sight velocity distribution of a uniformly expanding spherical shell.

If CSM interactions are responsible for this feature, the shock power must diminish over time to explain the \Ha disappearance. This implies a change in the density or structure of the local CSM environment at a radius of

\begin{equation}
\label{eqn:MLSize}
R\approx 1000 {\ \rm AU}\left(\frac{v_{\rm ej}}{6000\ \rm km/s}\right)\left(\frac{\Delta t}{300\ \rm d}\right)
\end{equation}
where the ejecta velocity, $v_{\rm ej}$, is estimated as the average of the early-time \Hb absorption velocity ($\sim8000$ km/s; \citealp{Tucker2024}) and the blueshifted [\ion{Ca}{2}] velocity ($\sim4000$ km/s) and $\Delta t=300\rm d$ is the approximate mid-point between the +211d spectrum from \citet{Ravi2025} and our latest spectrum at +370d. 

If this CSM is produced by a period of enhanced mass-loss from the progenitor in the years preceding the explosion, and we assume a typical RSG wind speed of $v_{\rm wind}\sim 10$ km/s \citep{windspeed}, then the enhanced mass-loss must begin 
\begin{equation}
\label{eqn:MLTime}
t_{\rm erupt}\approx500 {\ \rm yrs}\left(\frac{v_{\rm ej}}{6000\ \rm km/s}\right)\left(\frac{\Delta t}{300\ \rm d}\right)\left(\frac{v_{\rm wind}}{10\ \rm km/s}\right)^{-1}
\end{equation}
before explosion. If the progenitor was a yellow ($v_{\rm wind}\approx 100$ km/s; \citealp{RSGwind}) or blue ($v_{\rm wind}\approx 250$ km/s; \citealp{BSGWinds}) supergiant, the time scale could be much shorter.

\begin{table}[bt]
    \begin{center}       
    \begin{tabular}{ccccc}
    \hline
    \hline
    \rule[-1ex]{0pt}{3.5ex} Filter & Phase (d) & App Mag & Abs Mag\\
    \hline
    \rule[-1ex]{0pt}{3.5ex} $U$ & 760 & $>24.8$ & $>-9.3$\\
    \rule[-1ex]{0pt}{3.5ex} $B$ & 760 & $25.0\pm0.2$ & $-9.0\pm0.3$\\
    \rule[-1ex]{0pt}{3.5ex} $V$ & 760 & $>25.0$ & $>-9.1$\\
    \rule[-1ex]{0pt}{3.5ex} $R$ & 760 & $24.3\pm0.3$ & $-9.8\pm0.3$\\
    \rule[-1ex]{0pt}{3.5ex} $I$ & 760 & $>24.0$ & $>-10.0$\\
    \hline 
    \hline
    \end{tabular}
    \end{center}
    \caption{$UBVRI$ magnitudes of \name from the deep LBT photometry of the host galaxy. $3\sigma$ limits are reported for filters without a detection.} 
    \label{tab:SNMags}
\end{table} 

\begin{figure}
    \centering
    \includegraphics[width=\linewidth]{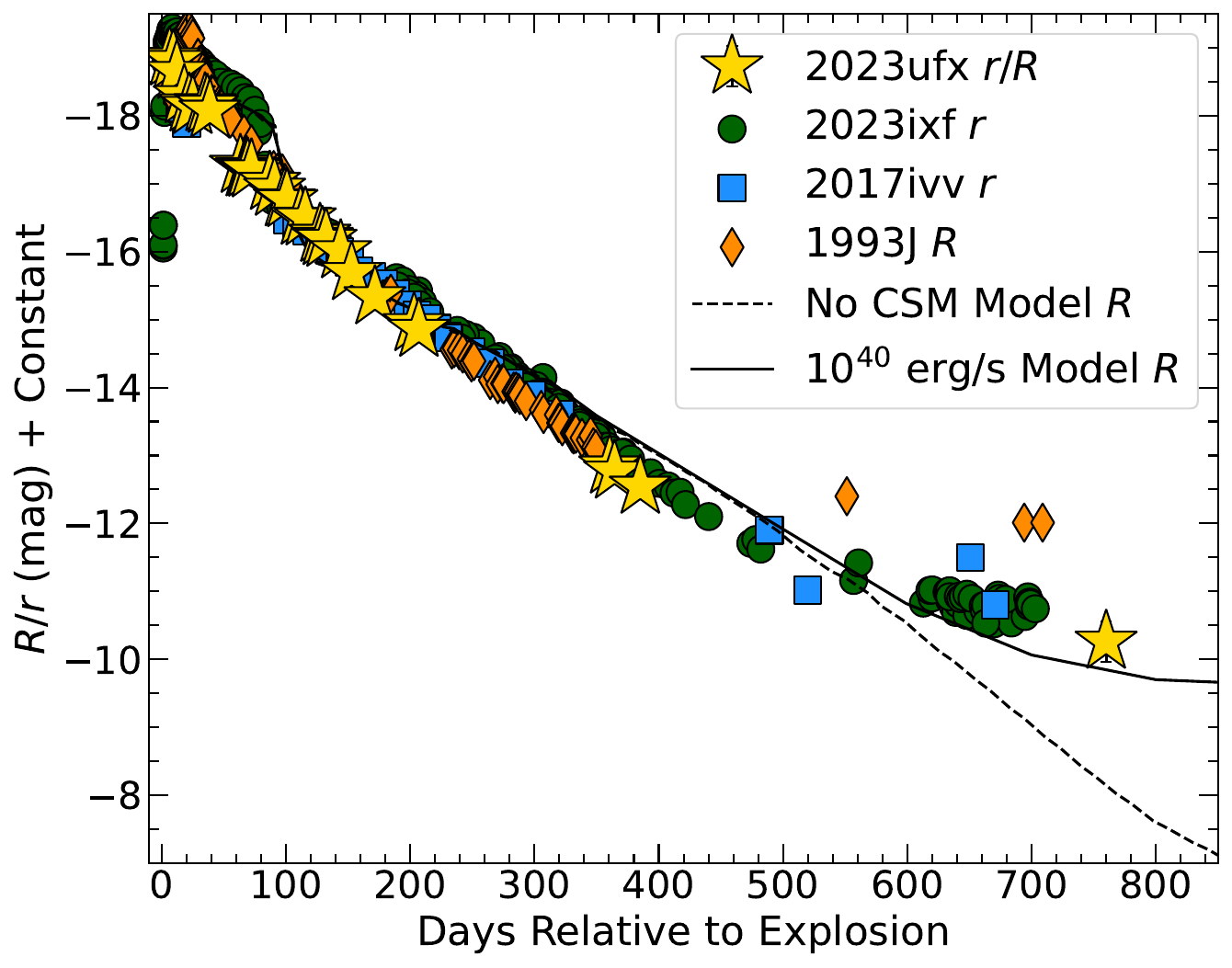}
    \caption{Comparison of the light curve of \name to other CCSNe and the models of \citet{DessartMorphDecay}. All CCSNe light curves are shifted to roughly align the radioactive decay tails during the early nebular phase. $r$-band photometry for \name comes from ZTF \citep{bellm2019} and MODS acquisition images, while the late-time (+760d) $R$-band photometry comes from our LBT observations of the host galaxy. Photometry of other CCSNe come from: SN 2017ivv \citealp{2017ivvGut}; SN 2023ixf \citealp{LaterixfPhot}; and SN 1993J \citealp{93JPhot, More93JPhot}}
    \label{fig:LcComp}
\end{figure}

As mentioned in \S\ref{sec:host}, the LBC images provide $\gtrsim3\sigma$ detections of \name 760 days after explosion in the $B$ and $R$ filters. We list magnitudes and $3\sigma$ detection limits for each LBC filter in Table \ref{tab:SNMags}. If the detections are in fact legitimate, then the measured magnitudes almost certainly require CSM interaction, as the radioactive-decay power must be lower. We compare to the ejecta-CSM interaction models of \citet{DessartMorphDecay} in Fig. \ref{fig:LcComp}, where the light curves of the observed CCSNe are shifted so that they roughly align with the radioactive decay tail during the earlier part of the nebular phase.

It is clear from the comparison that the LBC $R$-band photometry aligns more closely with the CSM interaction powered model. The CSM interaction powered model is associated with a typical RSG progenitor wind mass-loss rate of $10^{-6}$~\Msun/yr at a velocity of 10 km/s. Therefore, our comparison suggests a similar, or slightly higher, mass-loss rate to the model. \citet{RizzoSmith12CCSN} found that emission at this level is almost ubiquitous in Type II SN. Note that at this epoch, the broad, boxy \Ha emission has already significantly decreased in the nebular spectra of \name. Implications of this are further discussed in \S\ref{sec:Disc}.

\section{Discussion}\label{sec:Disc}

We present new observations of the extremely metal-poor SN II 2023ufx and its dwarf host. Combined with \citet{Tucker2024} and \citet{Ravi2025} we find that
\begin{enumerate}
  \item \name is one of the lowest metallicity CCSNe discovered thus far, with observations of the host yielding Z~$<0.1$~\Zsun.
  \item The low [\ion{Ca}{2}]/[\ion{O}{1}] ratio and comparisons to nebular spectra models imply a progenitor ZAMS mass of 25--35~\Msun.
  \item The nebular lines are anomalously broad and show structure not seen in any comparison CCSN. This seems to require a highly asymmetric explosion.
  \item The progenitor of \name appears to have had at least 2, and possibly 3, distinct mass-loss regimes in the last $\sim500$ yr before explosion.
\end{enumerate}

From Fig. \ref{fig:v_dist}, we roughly estimate the bulk velocities of the asymmetric [\ion{O}{1}] and [\ion{Ca}{2}] ejecta to be $\sim\pm3000$~km/s and $\sim 4100$~km/s respectively (Table \ref{tab:VelocityInfo}).  These ejecta velocities are comparable to the free-expansion ejecta velocities of intermediate-mass-elements in the asymmetric galactic supernova remnant Cassiopeia A ($2400-7100$~km/s; \citealp{CasAejecta}). They also align with SNR model predictions for the explosion of a progenitor that has been mostly stripped of its hydrogen envelope ($2000-7000$~km/s; \citealp{CasAModel}). However, note that these comparison velocities reflect deceleration by the reverse shock which has not yet occurred for \name at a phase of $\sim1$~yr.

Independent indicators of CSM interaction point towards at least 2, and possibly 3 distinct phases of mass-loss in the centuries leading up to explosion. First, \citet{Ravi2025} model the light curve of \name and find that the early times ($\lesssim10$~days) cannot be explained without introducing $\sim0.1$~\Msun of CSM, indicating an elevated mass loss of $3\times10^{-3}$~\Msun/yr in the 30 years before explosion assuming $v_{\rm wind}=10$ km/s. This CSM must be confined to within $\sim630$~AU.

Next, we find that the vanishing of the \Ha emission between the +211d and +370d spectra indicates a decrease in the density of the H-rich CSM as the ejecta expands. The inferred radius at which the \Ha emission disappears is $\sim1000$ AU (Eq. \ref{eqn:MLSize}), relatively close to the extent inferred by \citet{Ravi2025} ($\sim630$ AU). This could mean that the boxy \Ha profile is tracing the tail end of the CSM interaction powering the early light curve. Then, the $\sim 1000\times$ weaker CSM interaction seen at $\sim2$ yr in the LBT imaging is tracing the `normal' stellar wind. \citet{Tucker2024} point out the issue of `missing' UV flux from the increased CSM interaction, but that may be reconcilable with more sophisticated radiative-transfer models.

To summarize the mass-loss history, we estimate that the progenitor of \name had a fairly typical mass-loss rate of $\sim10^{-6}-10^{-5}$~\Msun/yr that increased by $\sim100-1000\times$ in the centuries to millennia leading up to explosion.

SN~1993J offers an interesting comparison to \name. Their nebular spectra are broadly similar, showing boxy \Ha (Fig. \ref{fig:CompSNe}) and similar emission-line ratios (Fig. \ref{fig:TimeEvolve}). Both progenitors had thin H envelopes and their early light curves show a prominent shock-cooling peak, requiring CSM in the vicinity of the progenitor at the moment of explosion (M$_{\rm CSM}\sim0.06$~\Msun for SN 1993J; \citealp{93JCSM_Mass}). While there are quantitative differences with SN 1993J, which had a thinner H envelope ($\sim 0.2$~\Msun; \citealp{93JHenv2, 93JHenv3, 93JHenv1}) and a lower progenitor mass (13--22~\Msun; \citealp{1993JProM}), it shares many qualities with \name.

The agreement with SN~1993J might imply that metallicity does not play a direct, overarching role in the properties of \name. SN~1993J exploded in the nearby and metal-rich (Z$\sim2$~\Zsun; \citealp{93JMetal}) galaxy M81, strongly disfavoring a metal-poor progenitor. Instead, the properties of SN 1993J are often attributed to binarity \citep{93JBinar1, 93JBinar2}, and there is a candidate surviving companion \citep{93JComp, 93JCompHST}. 

Thus, some of the peculiar properties of \name may instead trace the configuration of the pre-explosion binary system. Massive stars prefer massive companions, and $\sim$ 70\% will experience some phase of interaction or mass-transfer \citep{BigStarBinary} during their (co-)evolution. Mass transfer can also produce a wide variety of local CSM environments at the moment of explosion depending on the configuration of the pre-explosion binary \citep{BinaryCliffs, BinaryCSM1}, perhaps explaining the sharp decrease in \Ha emission seen in \name. 

More evidence for the progenitor of \name residing in a binary system comes from comparing to the models of \citet{sukhbold2016}. They find that a solar metallicity progenitor with a ZAMS mass of 35~\Msun ends its life with an envelope mass of M$_{\rm env}\sim1$~\Msun. We see a similar M$_{\rm env}$ for \name at only $\sim0.1$~\Zsun. If we adopt \Mdot $\propto$ Z$^{0.7}$ from \citet{vink2001}, we would expect the progenitor of \name to have a mass-loss rate of only 20\% that of an identical progenitor evolving at solar metallicity. Therefore, we would expect a much greater envelope mass than the estimates of \citet{Tucker2024} and \citet{Ravi2025} if the progenitor evolved in a single star system.

\name is one entry in the small, but growing, list of nearby CCSNe with Z~$\lesssim0.1$~\Zsun. Metallicity must have some effect on CCSN explosions, but the magnitude of the effect(s) and its observational manifestations remain unclear. Lower metallicities decrease mass-loss rates, which could produce faster-spinning progenitors with more massive H envelopes at collapse. However, wind-driven mass loss may be superseded by binary interaction once the metallicity is sufficiently low. The local metal-poor CCSN rate is inherently limited by the local space density of dwarf galaxies, highlighting the importance of well-studied events like \name for anchoring theoretical models. Finding and characterizing more examples is crucial to better understand how CCSNe and, by extension, galactic feedback and evolution, have evolved over cosmic time.

\section*{Acknowledgments}

We thank David Jones for assistance with \textsc{Blast}.

We would like to thank Paul Crowther, James Trussler, Annika Deutsch, and Raphael Baer-Way for insightful conversations. 

EJ and CSK are supported by NSF grants AST-
2307385 and 2407206.

WBH acknowledges support from the National Science Foundation Graduate Research Fellowship Program under Grant No. 2236415.


This work is based on observations made with the Large Binocular Telescope. The LBT is an international collaboration among institutions in the United States, Italy, and Germany. LBT Corporation partners are: The University of Arizona on behalf of the Arizona Board of Regents; Istituto Nazionale di Astrofisica, Italy; LBT Beteiligungsgesellschaft, Germany, representing the Max-Planck Society, The Leibniz Institute for Astrophysics Potsdam, and Heidelberg University; The Ohio State University, representing OSU, University of Notre Dame, University of Minnesota and University of Virginia.

This paper used data obtained with the MODS spectrographs built with
funding from NSF grant AST-9987045 and the NSF Telescope System
Instrumentation Program (TSIP), with additional funds from the Ohio
Board of Regents and the Ohio State University Office of Research.

Observations have benefited from the use of ALTA Center (alta.arcetri.inaf.it) forecasts performed with the Astro-Meso-Nh model. Initialization data of the ALTA automatic forecast system come from the General Circulation Model (HRES) of the European Centre for Medium Range Weather Forecasts.

This research used the facilities of the Italian Center for Astronomical Archive (IA2) operated by INAF at the Astronomical Observatory of Trieste.

%

\vspace{5mm}
\facilities{Keck:II (KCWI), LBT (MODS, LBC, LUCI)}


\software{astropy \citep{astropy}, matplotlib \citep{matplotlib}, lmfit \citep{lmfit}, spectres \citep{spectres}, numpy \citep{numpy}, pandas \citep{pandas}, extinction \citep{extinction}, scipy \citep{scipy}, astro-scrappy \citep{astroscrappy}, Bagpipes \citep{Bagpipes}, Blast \citep{Blast}, photutils \citep{photutils}, astroquery \citep{astroquery}, HiPS \citep{HiPS}, dynesty \citep{Dynesty}, Prospector \citep{Prospector}, SVO Filter Profile Service \citep{SVOFilter}, python-fsps \citep{python-fsps}, SBI++ \citep{sbipp}}










\bibliography{ref}{}
\bibliographystyle{aasjournal}



\end{document}